\documentclass[aps,prd]{revtex4}
\usepackage{graphicx}
\begin{document}
\begin{flushright}{OITS 716}\\
April 2002
\end{flushright}
\voffset 2cm
 \vspace*{1cm}
\title{Inclusive Distributions for Hadronic
Collisions in the Valon-Recombination Model}

\author{Rudolph C. Hwa$^1$ and C.\ B.\ Yang$^{1,2}$}

\affiliation{$^1$Institute of Theoretical Science and Department of
Physics\\ University of Oregon, Eugene, OR 97403-5203, USA\\
\bigskip
$^2$Institute of Particle Physics, Hua-Zhong Normal
University, Wuhan 430079, P.\ R.\ China}

\begin{abstract}
Inclusive distributions of soft production of mesons in hadronic
collisions are calculated in the valon-recombination model. The
new determination of the valon distributions from hard scattering
data makes possible a tightly interrelated treatment of pion and
kaon production in the fragmentation regions of proton, pion and
kaon. Only one free parameter is used in the determination of the
valon distribution in the kaon. No other adjustable parameter is
needed to fit the $x$-distributions of the data on inclusive cross
sections, except for the normalizations since the data are at
fixed $p_T$. The success of the model in reproducing seven
inclusive distributions suggests that there are two important
mechanisms at work in soft production. One is that the structure
of the hadron that fragments is highly relevant. The other is that
the produced particles are formed by the recombination of quarks
and antiquarks. These two aspects about hadrons in soft processes
can be well described in the framework of the valon-recombination
model.
\end{abstract}

\maketitle

\section{Introduction}

Particle production at low $p_T$ in hadronic collisions has
always been a challenge for theoretical models to describe, since,
on the one hand, it is a process that is non-perturbative, while,
on the other hand, meaningful modeling can only be done in the
framework of quarks and gluons.  The subject has a long history,
some early reviews of which can be found in Refs.\ \cite{vw,fk}.
Since soft processes, as they are called, cannot be treated by
perturbative QCD, they have not been given the degree of
rigorous scrutiny that have been accorded hard processes.
Among the models that are constructed for soft processes, there
are basically only two types:  string models that make use of
fragmentation \cite{dpm,lu} and parton models that are based
on recombination \cite{dh,val}.  This paper treats the
modernization of the latter.  What is new is that the parton
distributions of the proton have recently been calculated by CTEQ in
pQCD over wide ranges of $x$ and $Q^2$, fitting a large
collection of experimental data \cite{cteq}.  From such parton
distributions the structure of the proton in terms of the valons
\cite{hz} can be more precisely determined \cite{hy}.  With the
new parameterization of the valon model now available, it is
possible to revisit the problem of hadron production in soft
processes and calculate the inclusive distributions of the
produced particles without adjustable parameters.  Good
agreement with the low-$p_T$ experimental data can
give definitive support to the recombination model.

The kinematical region in which we focus our attention is the
projectile fragmentation region, roughly $x > 0.2$.  In the
central region the structure of the projectile hadron is less
important. But in the fragmentation region it is known as early as
the mid-70's that the inclusive distribution of the produced pions
is closely related to the structure function of the proton, as
observed by Ochs \cite{wo}.  The recombination model \cite{dh} is
a realization of that observation, and the valon model is a
self-consistent formulation of the unification of hadron structure
and recombination probability \cite{val}.  In short, our view is
that, while the fragmentation of strings may be suitable for the
central region, the recombination of partons is more relevant for
the fragmentation region.

It has been regarded as a striking confirmation of the dual parton
model (DPM) \cite{dpm} to reproduce the charge distribution of the
produced particles in $\pi^+p$ collisions; the forward-backward
asymmetry \cite{fig} is presented as evidence for the two-chain
diquark fragmentation of the proton.  While the qualitative
agreement with data gives support to the two-chain mechanism
of DPM as opposed to the single-chain mechanism of the LUND
model, the data (not cited in \cite{dpm}) were old and
inaccurate and the theoretical calculations were crude.  The data
that we shall compare with are detailed and more precise
\cite{b}.  Moreover, we shall consider a large number of
fragmentation processes:  $p \rightarrow \pi^{\pm}$,
$\pi^{\pm} \rightarrow \pi^{\mp}$
, $K^+ \rightarrow \pi^{\pm}$, and $\pi^+ \rightarrow
K^{\pm}$, and the results of our calculations can all be
compared with existing data.  To our knowledge such data have
never been used before to confront model calculations.  That the
valon-recombination model is able to reproduce those data, as we
shall show in this paper,  should therefore be regarded as
meeting a higher demand than fitting the charge asymmetry of
$\pi^+p$ collisions discussed in \cite{dpm}.

Another reason for revisiting the soft processes in hadronic
collisions is that the recent development in heavy-ion collisions
at high energies presents some urgency to understand better the
basic processes of particle production at a more fundamental
level.  To understand the formation of dense matter, it is
necessary to understand first baryon stopping and pionization
in $pA$ collisions.  We have recently done a comprehensive
treatment of the problem of momentum degradation in $pA$
collisions in the framework of the valon model \cite{hy2}.  The
emphasis is on the nuclear effects on the projectile.  The
parameters used for the valon and parton distribution functions
have not been put to test on the hadronic collisions.  Since those
distributions have been updated even more recently \cite{hy}, it
is necessary to focus on the elementary processes, not only of
$pp$, but also of meson-proton collisions, in order to check the
valon-recombination model (VRM) to a degree never attempted
before.  It is toward that end that we apply the VRM to the soft
processes in this paper.

\section{Valon and Parton Distribution Functions}

The valon model describes the hadron structure relevant for
multiparticle production.  The momentum distributions of the
valons can be determined from the parton distributions at low
$Q^2$, as posted by CTEQ4LQ \cite{cteq4}.  That was done very
recently \cite{hy}, resulting in excellent fits of the $u$ and $d$
quark distributions.  The parameters describing the valon
distributions turn out to be quite different from the old values
based on imprecise muon and neutrino data of the 70's
\cite{hz}, although the formalism of the model remains the
same.  Furthermore, the assumption of the symmetric sea used
previously has been lifted so that the parton distributions in the
valons are now also very different.  Here we give a summary of
the valon and parton distributions functions, the details of
which can be found in \cite{hy}.

In a proton the 3-valon distribution is
\begin{eqnarray}
G_{UUD}(y_1, y_2, y_3) = g_p \, (y_1
y_2)^{\alpha}y^{\beta}_3
\, \delta (y_1 + y_2 + y_3 -1) ,
\label{2.1}
\end{eqnarray}
where $y_i$ is the momentum fraction of the {\it i}th valon ($y$
being never used for rapidity in this paper), and
\begin{eqnarray}
g_p = \left[B (\alpha + 1, \beta +1) B(\alpha + 1, \alpha +\beta
+2)\right]^{-1},
\label{2.2}
\end{eqnarray}
$B(m,n)$ being the beta function.  The single valon
distributions are obtained by  integration
\begin{eqnarray}
G_U(y) = \int dy_2 \int dy_3 G_{UUD} (y, y_2, y_3) = g_p B
(\alpha + 1,
\beta +1) y^{\alpha} (1 - y)^{\alpha +\beta +1},
\label{2.3}
\end{eqnarray}
\begin{eqnarray}
G_D(y) = \int dy_1 \int dy_2 G_{UUD} (y_1, y_2, y)= g_p B
(\alpha + 1, \alpha +1) y^{\beta} (1 - y)^{2\alpha +1}.
\label{2.4}
\end{eqnarray}
The new values of $\alpha$ and $\beta$ are found to be
\cite{hy}
\begin{eqnarray}
\alpha = 1.76 \hspace{.5in}\mbox{and} \hspace{.5in}\beta =
1.05 ,
\label{2.5}
\end{eqnarray}
which are significantly different from those used in Refs.\
\cite{val,hy2} due to various theoretical assumptions and limited
experimental data.

Since each valon contains one and only one valence quark of its
own flavor, the valence quark distributions are convolutions of
the relevant valon distributions with the valence (non-singlet)
quark distributions in the valons.  In terms of moments we have
simple products
\begin{eqnarray}
\tilde{u}_v  (n) =  2\tilde{G}_U(n) \tilde{K}_{NS} (n),
\label{2.6}
\end{eqnarray}
\begin{eqnarray}
\tilde{d}_v  (n) =  \tilde{G}_D(n) \tilde{K}_{NS} (n),
\label{2.7}
\end{eqnarray}
where
\begin{eqnarray}
\tilde{G}_{U, D} (n) = \int^1_0 dy y^{n-1} G_{U, D} (y),
\label{2.8}
\end{eqnarray}
and similarly for $\tilde{q}_v(n)$ in terms of $q_v(x)$ where
$q$ stands for either $u$ or $d$.  The value of $Q^2$ is set at
$1 (GeV/c)^2$ for application of the model to low-$p_T$
processes and will not be exhibited explicitly.  From Eqs.\
(\ref{2.3}) and (\ref{2.4}) we have
\begin{eqnarray}
\tilde{G}_U(n) = B (\alpha + n, \alpha + \beta + 2)/B(\alpha + 1,
\alpha + \beta +2)
\label{2.9}
\end{eqnarray}
\begin{eqnarray}
\tilde{G}_D(n) = B (\beta + n, 2 \alpha + 2)/B(\beta + 1,
2 \alpha + 2) .
\label{2.10}
\end{eqnarray}
The $n$ dependence of $\tilde{K}_{NS}(n)$ is determined in
\cite{hy} and fitted by the following parameterization
\begin{eqnarray}
\tilde{K}_{NS}(n) = \exp \left(- \sum^3_{j=0} c_j u^j\right),
\qquad \qquad  u = \ln (n-1),
\label{2.11}
\end{eqnarray}
where $c_j = 0.753$, $0.401$, $0.0962$, and $0.0555$, for $j =
0$, $1$, $2$, $3$, respectively.

For the sea quark distributions in the valons we have found that
the SU(2) symmetry has to be broken to fit the CTEQ4 data, and
that the favored quark distributions ($u$ in $U$ and $d$ in
$D$) are suppressed relative to the unfavored quark
distributions ($u$ in $D$ and $d$ in $U$), consistent with Pauli
blocking.  They are denoted, respectively, by $L_f(z)$ and
$L_u(z)$, which are different from the $s$-quark distribution
$L_s(z)$ and the gluon distribution $L_g(z)$.  The distributions
of these partons in the proton are given, in terms of the
moments, simply by sums of products
\begin{eqnarray}
\tilde{\bar{u}} = 2 \tilde{G}_U \tilde{L}_f + \tilde{G}_D
\tilde{L}_u ,
\label{2.12}
\end{eqnarray}
\begin{eqnarray}
\tilde{\bar{d}} = \tilde{G}_D \tilde{L}_f + 2
\tilde{G}_U \tilde{L}_u ,
\label{2.13}
\end{eqnarray}
\begin{eqnarray}
\tilde{s} = \left(2 \tilde{G}_U + \tilde{G}_D \right)\tilde{L}_s ,
\label{2.14}
\end{eqnarray}
\begin{eqnarray}
\tilde{g} = \left(2 \tilde{G}_U + \tilde{G}_D \right)\tilde{L}_g
\label{2.15}
\end{eqnarray}
where the dependences on $n$ have been omitted.  The
functions $\tilde{L}_i(n)$ are parameterized as follows
\cite{hy}
\begin{eqnarray}
\ln  \, \tilde{L}_i(n) = - \sum^3_{j = 0} b^{(i)}_j u^j, \qquad u =
\ln (n-1)
\label{2.15a}
\end{eqnarray}
where $b^{(i)}_j$ are given in Table I.
\begin{table}[h]
\begin{center}
\caption{Coefficients in Eq.\ (\ref{2.15a})}
\begin{tabular}{|c|c|c|c|c|}\hline
$i$&$b^{(i)}_0$&$b^{(i)}_1$&$b^{(i)}_2$&$b^{(i)}_3$\\\hline
$f$&4.12&2.2&0.2&0.18\\\hline
$u$&3.07&1.5&0.08&0.05\\\hline
$s$&4.21&1.6&0.1&0.02\\\hline
$g$&0.98&1.0&0.05&0\\\hline\hline
\end{tabular}
\end{center}
\end{table}

For meson-initiated reactions we need the valon distributions in
mesons, which can be determined from the quark distributions
in mesons on the basis of the universality of the quark
distributions in valons.  From experimental data on Drell-Yan
and prompt photon production in $\pi^{\pm}N$ collisions, the
parton distributions  in the pion have been determined in Ref.\
\cite{sut}, using the parameterization
\begin{eqnarray}
xq_v(x) = A_v x^{\alpha^\prime}(1-x)^{\beta^\prime}
\label{2.16}
\end{eqnarray}
for the valence quarks with $\alpha^\prime = 0.64\pm 0.03$ and
$\beta^\prime = 1.11\pm 0.04$.  In the valon model for the pion
that distribution is related to the valon distribution by
\begin{eqnarray}
xq_v(x) = \int^1_0 dy_1 \int^{1-y_1}_0 dy_2 G^{\pi}\left(
y_1,y_2\right) K_{NS} \left(x/y_1\right),
\label{2.17}
\end{eqnarray}
where $K_{NS}(z)$ is identified with the same in the proton
problem.  Since the two valons in pion are symmetrical, we use
the valon distribution
\begin{eqnarray}
G^{\pi}\left(y_1,y_2\right) =
g_{\pi}\left(y_1y_2\right)^{\gamma} \delta\left(y_1 + y_2
- 1\right)
\label{2.18}
\end{eqnarray}
where $g_{\pi} = 1/B(\gamma +1, \gamma +1)$.  We convert
Eq.\ (\ref{2.17}) to the moment form
\begin{equation}
\tilde q_v(n) = \tilde G^\pi(n)\, \tilde K_{NS}(n),   \label{2.18a}
\end{equation}
where $\tilde G^\pi(n)$ involves only one parameter $\gamma$,
and $\tilde K_{NS}(n)$ is the same as in Eq.\ (\ref{2.11}) on the basis
of the universality of the valon structure, independent of the host
hadron.  The corresponding moment of Eq.\ (\ref{2.16}) can
trivially be calculated. Both are shown in Fig.\ 1, where a good
agreement between the two is achieved by the  choice
\begin{equation}
\gamma=0.     \label{2.18b}
\end{equation}
This value is also consistent with the one used in Ref.\ \cite{val}.
 Thus $g_{\pi} =1$, and the single valon distribution in
a pion is simply
\begin{eqnarray}
G^{\pi} (y) = 1 .
\label{2.19}
\end{eqnarray}
The flat distribution of the valons in pion is a result of the fact
that the constituent quarks are much more massive than the pion
so they are tightly bound, resulting in large uncertainty in the
valon momentum fraction.

The situation with the $K$ meson is somewhat different, since
the kaon mass is higher, and the constituent quarks have
unequal masses.  The valon distribution has a form similar to that
of the pion in Eq.\ (\ref{2.18})
\begin{eqnarray}
G^K \left(y_1,y_2\right) = g_Ky^a_1 y^b_2 \delta\left(y_1+y_2
- 1\right)
\label{2.20}
\end{eqnarray}
where $g_K = 1/B (a + 1, b + 1)$.  The average momentum
fractions of the two valons are
\begin{eqnarray}
\overline{y}_{1,2} = \int dy_1 dy_2 y_{1,2} G^K
\left(y_1, y_2\right)
\label{2.21}
\end{eqnarray}
so their ratio is
\begin{eqnarray}
\overline{y}_1/\overline{y}_2 = (a + 1)/(b + 1) .
\label{2.22}
\end{eqnarray}
Since the average velocities of the two valons are the same as that
of the host kaon, their momenta should be proportional to their
masses, i.\ e., $\overline{y}_1/\overline{y}_2 \approx
m_U/m_S$.  Taking the constituent quark masses of $u$- and
$s$-types to be in the ratio 2:3, we get from Eq.\ (\ref{2.22})
\begin{eqnarray}
b = (3 a + 1)/2 .
\label{2.23}
\end{eqnarray}
Thus we are left with one unknown parameter in Eq.\
(\ref{2.20}).  Since the parton distributions in kaon are not
known, we cannot determine that parameter as in the pion case.
It will, however, be determined by the soft production data of
kaon-initiated collisions later in Sec.\ 5.

Knowing the valon distributions in mesons results in our
knowledge of the recombination functions for the formation of
the same mesons.  Since we define the recombination function in
the invariant phase space, we have from (\ref{2.18}) and
(\ref{2.20})
\begin{eqnarray}
R_{\pi}\left(x_1, x_2, x\right) = {x_1  x_2 \over x^2} \delta
\left({x_1  \over x} + {x_2 \over x} - 1\right)
\label{2.24}
\end{eqnarray}
\begin{eqnarray}
R_K \left(x_1, x_2, x\right) = g_K \left({x_1  \over
x}\right)^{a + 1}\left(x_2 \over x \right)^{b + 1}
\delta
\left({x_1  \over x} + {x_2 \over x} - 1\right)
\label{2.25}
\end{eqnarray}
with $a$ and $b$ being constrained by (\ref{2.23}).  In this
paper we do not consider nucleon production, so $R_p$ is not
needed (see \cite{hy2}).

\section{Proton Fragmentation}

In a $pp$ collision the soft production process in the
fragmentation region is treated in the VRM as one in which the
proton bag is broken by the collision and the valons become
clusters of partons.  The central idea of the model is that the
probability for detecting a meson at large $x$ is higher for a
$q$ and a $\bar{q}$ at lower $x_i$ to recombine than for a $q$
or a diquark at high $x^{\prime}$ to fragment.  The
distributions of $q(x_1)$ and $\bar{q}(x_2)$ depend on the
hadron structure.  Under the assumption that the collision
process does not significantly perturb the parton distributions
outside the central interaction region (i.e., $x_i \geq 0.1$), the
valon model can provide a sensible link between hadron
structure and parton distributions.  Since glueballs have never
been seen, the gluons hadronize by first converting to
$q\bar{q}$ pairs, thereby enhancing the sea.  Downstream the
quarks and antiquarks dress themselves and become the valons
to be recombined in forming  the produced particles.  Since the
dressing process does not change the total momentum of a
quark, the meson momentum $x$ is simply the sum of $x_1$
and $x_2$ of the $q$ and $\bar{q}$, and the probability of
hadronization is determined by an integration of the enhanced
$q(x_1)$ and $\bar{q}(x_2)$ weighted by the recombination
function.  This is an $s$-channel description of the
fragmentation process, the initial formulation of which is given
in Ref.\ \cite{val}.  What we have now are the new distributions
\cite{hy} and the more extensive low-$p_T$ data \cite{b} not
available to be considered in \cite{val}.

Quantitatively, the invariant distribution for pion production in
the proton fragmentation region is
\begin{eqnarray}
x{dN_{\pi} \over dx}
 = H_{\pi}(x) = \int {dx_1 \over x_1} {dx_2 \over
x_2}F_{\pi}\left(x_1, x_2 \right) R_{\pi}\left(x_1, x_2 ,x\right)
\label{3.1}
\end{eqnarray}
where $F_{\pi}\left(x_1, x_2 \right)$ is the invariant
distribution for $q$ at $x_1$
 and $\bar{q}$ at $x_2$, and is a convolution of the valon
distribution in a proton with the quark distributions in valons,
whose sea quarks are enhanced.  In Fig.\ 2 we show the
subprocesses giving $u$ and $\bar{d}$ quarks by schematic
diagrams, which represent
\begin{eqnarray}
F_{\pi^+}\left(x_1, x_2 \right) = \int dy_1 dy_2 \left[ G_{UU}
\left(y_1, y_2 \right) K \left(x_1/y_1 \right)
L^{\prime}_u\left(x_2/y_2 \right)\right.\nonumber\\
+ 2G_{UD}
\left(y_1, y_2 \right) K\left(x_1/y_1 \right)
L^{\prime}_u\left(x_2/y_2 \right)\nonumber\\
+ \left.2G_{UD} \left(y_1, y_2 \right) L^{\prime}_u\left(x_1/y_1
\right)L^{\prime}_u\left(x_2/y_2 \right)\right]\nonumber\\
+\int dy \left[ 2G_U (y) \left\{ K\left(x_1 \over y
\right)L^{\prime}_u\left(x_2 \over y - x_1
\right)\right\}_{12}\right.  \nonumber\\
\left. + G_D (y) \left\{ L^{\prime}_u\left(x_1 \over y
\right)L^{\prime}_u\left(x_2 \over y - x_1
\right)\right\}_{12} \right] ,
\label{3.2}
\end{eqnarray}
where
\begin{eqnarray}
K(z) = K_{NS} (z) + L^{\prime}_f (z) .
\label{3.3}
\end{eqnarray}
$L^{\prime}_f (z)$ and $L^{\prime}_u (z)$ are favored and
unfavored sea quark distributions in a valon, enhanced by gluon
conversion, and will be discussed below.  In Eq.\ (\ref{3.2}) the
symbol $\{\dots \}_{12}$ implies symmetrization in $x_1$ and
$x_2$.  Equation (\ref{3.2}) can appear simpler in moment
form, for which we define
\begin{eqnarray}
\tilde{F}_{\pi^+}\left(n_1, n_2 \right) = \int^1_0 dx_1
\int^{1-x_1}_0 dx_2 x_1^{n_1-2}
x_2^{n_2-2}F_{\pi^+}\left(x_1, x_2 \right) ,
\label{3.4}
\end{eqnarray}
so that (\ref{3.2}) becomes
\begin{eqnarray}
\tilde{F}_{\pi^+}\left(n_1, n_2 \right) =  \left[
\tilde{G}_{UU}
\left(n_1, n_2 \right) + \left.2\tilde{G}_{UD}\left(n_1, n_2
\right)\right] \tilde{K}
\left(n_1 \right)
\tilde{L}^{\prime}_u\left(n_2\right)\right.\nonumber\\
+ 2\tilde{G}_{UD}
\left(n_1, n_2 \right)
\tilde{L}^{\prime}_u\left(n_1\right)
\tilde{L}^{\prime}_u\left(n_2\right)\nonumber\\ +
2\tilde{G}_U \left(n_1+ n_2 - 1 \right)\left\{\tilde{K}
\left(n_1, n_2 \right)
\tilde{L}^{\prime}_u\left(n_2\right) \right\}_{12}
\nonumber\\
 + \tilde{G}_D \left(n_1+ n_2 - 1 \right) \left\{
\tilde{L}^{\prime}_u \left(n_1,
n_2 \right) \tilde{L}^{\prime}_u \left(n_2\right)\right\}_{12}
\label{3.5}
\end{eqnarray}
where
\begin{eqnarray}
\tilde{K}
\left(n_1, n_2 \right) = \int^1_0 dz z^{n_1 - 2} (1 - z)^{n_2 -
1} K(z)
\label{3.6}
\end{eqnarray}
and similarly for $\tilde{L}^{\prime}_u\left(n_1, n_2 \right)$.
For $\pi^-$ we have
\begin{eqnarray}
\tilde{F}_{\pi^-}\left(n_1, n_2 \right) =  \left[
\tilde{G}_{UU}
\left(n_1, n_2 \right) +  2\tilde{G}_{UD}\left(n_1, n_2
\right) \right] \tilde{L}^{\prime}_u\left(n_1\right)
\tilde{L}^{\prime}_u\left(n_2\right)\nonumber\\
+ 2\tilde{G}_{UD}
\left(n_1, n_2 \right)
\tilde{L}^{\prime}_u\left(n_1\right)
\tilde{K}\left(n_2\right)\nonumber\\ +
2\tilde{G}_U \left(n_1+ n_2 -1\right)
\left\{\tilde{L}^{\prime}_u \left(n_1, n_2 \right)
\tilde{L}^{\prime}_u\left(n_2\right) \right\}_{12}
\nonumber\\
 + \tilde{G}_D \left(n_1 + n_2 - 1 \right) \left\{
\tilde{K} \left(n_1,
n_2 \right) \tilde{L}^{\prime}_u \left(n_2\right)\right\}_{12} .
\label{3.7}
\end{eqnarray}

To use these in Eq.\ (\ref{3.1}) we note that the moments of
$H^{\prime}_{\pi} (x) = x^3H_{\pi}(x)$ is, using Eq.\
(\ref{2.24}),
\begin{eqnarray}
\tilde{H}^{\prime}_{\pi} (n)= \int^1_0 dx x^{n -
2}H^{\prime}_{\pi} (x)
= \int^1_0 dx_1 \int^{1-x_1}_0 dx_2 F_{\pi}\left(x_1,
x_2\right) \left(x_1 + x_2\right)^n \nonumber\\
= \sum_{[n_i]} {n! \over n_1! n_2!} \tilde{F}_{\pi}\left(n_1+2,
n_2+2\right) ,
\label{3.8}
\end{eqnarray}
where the sum over $n_1$ and $n_2$ is restricted by the
constraint $n_1 + n_2 = n$.  The double moments of the 2-valon
distributions are
\begin{eqnarray}
\tilde{G}_{UU}
\left(n_1, n_2 \right) = \int dy_1
dy_2 y_1^{n_1-1} y_2^{n_2-1}G_{UUD}\left(y_1, y_2,
y_3\right)
\nonumber\\
=g_p \int ^1_0 dy_1 \int ^{1-y_1}_0 dy_2 y_1^{n_1+ \alpha
-1}  y_2^{n_2+ \alpha -1} \left(1 -y_1-
y_2\right)^{\beta} \nonumber\\
=g_p B\left(n_1+ \alpha, n_2+ \alpha + \beta
+1\right)  B\left(n_2+ \alpha,  \beta
+1\right)  ,
\label{3.8a}
\end{eqnarray}
\begin{eqnarray}
\tilde{G}_{UD}
\left(n_1, n_2 \right) = \int dy_1
dy_3 y_1^{n_1-1} y_3^{n_2-1}G_{UUD}\left(y_1, y_2,
y_3\right)
\nonumber\\
=g_p B\left(n_1+ \alpha, n_2+ \alpha + \beta
+1\right)  B\left(n_2+ \beta, \alpha
+1\right)  ,
\label{3.8b}
\end{eqnarray}
where $g_p$ is given by Eq.\ (\ref{2.2}).

For the enhanced sea quark distributions, $L^{\prime}_f (z)$
and $L^{\prime}_u (z)$, we recall that the quiescent sea as
probed by electroweak interaction is described by $L_f (z)$ and
$L_u(z)$, given in Ref.\ \cite{hy}.  For either a $U$ or $D$
valon, the momentum fractions of all its partons add up to one,
so we have a constraint on the $n = 2$ moments
\begin{eqnarray}
\tilde{K}_{NS}(2) + 2 \left[\tilde{L}_f(2) + \tilde{L}_u(2) \right] +
2 \tilde{L}_s(2) + \tilde{L}_g(2) = 1 .
\label{3.9}
\end{eqnarray}
The various terms above correspond to (for  $U$ valon, say) $u$
valence, $u\bar{u}$ sea, $d\bar{d}$ sea, $s\bar{s}$ sea and
gluons.  For hadronization we consider the enhanced sea where
the gluons are completely converted into the $u\bar{u}$ and
$d\bar{d}$ sectors, a scenario which we refer to as the saturated
sea.  Note that we do not let the gluons be converted to the
$s\bar{s}$ sector due to the higher $s$-quark mass.  Such a
restriction should be relaxed in the case of nuclear collisions
because of substantial Pauli blocking in the non-strange sectors.
Thus we write
\begin{eqnarray}
\tilde{K}_{NS}(2) + 2
\left[\tilde{L}^{\prime}_f(2) + \tilde{L}^{\prime}_u(2)
\right] + 2 \tilde{L}_s(2)  = 1 .
\label{3.10}
\end{eqnarray}
From these two equations follows
\begin{eqnarray}
\tilde{L}_g(2) =2 \left[\tilde{L}^{\prime}_f(2) +
\tilde{L}^{\prime}_u(2) - \tilde{L}_f(2) - \tilde{L}_u(2)
\right]  .
\label{3.11}
\end{eqnarray}
Assuming that the enhancement factor $f_q$ is the same for
favored and unfavored quarks, i.e., $\tilde{L}^{\prime}_{f,u}(2)
= f_q \tilde{L}_{f,u}(2)$,  we get
\begin{eqnarray}
f_q = 1 + {\tilde{L}_g(2)\over2 \left[\tilde{L}_f(2) +
\tilde{L}_u(2)\right]} .
\label{3.12}
\end{eqnarray}
From Table I,  the values of $\ln \tilde{L}_i(2)$ are found to be
$-4.12$, $-3.07$ and $-0.98$, for $i = f$, $u$, and $g$,
respectively.  One thus obtains
\begin{eqnarray}
 f_q = 3.99  .
\label{3.13}
\end{eqnarray}
We assume that this enhancement factor applies uniformly at all
$z$ so that we have
\begin{eqnarray}
L^{\prime}_{f,u}(z)
= f_q L_{f,u}(z) , \quad \mbox{or} \quad
\tilde{L}^{\prime}_{f,u}(n) = f_q \tilde{L}_{f,u}(n) .
\label{3.14}
\end{eqnarray}
This assumption is made mainly for the sake of simplicity and to
avoid introducing undetermined parameters.  It cannot be
expected to be valid at very low $z$ where the density is high,
but that is outside the region of applicability of VRM anyway.

The two last quantities in Eqs.\ (\ref{3.5}) and (\ref{3.7}) that
remain to be specified are $\tilde{K}(n_1, n_2)$ and
$\tilde{L}^{\prime}_{u}(n_1, n_2)$.  They both are defined by
integrals of the typed given in (\ref{3.6}) in terms of
$K_{NS}(z)$, $L^{\prime}_f(z)$ and $L^{\prime}_u(z)$, the
latter two being proportional to $L_f(z)$ and $L_u(z)$.
Since the single moments $\tilde{K}_{NS}(n)$, $\tilde{L}_f(n)$
and $\tilde{L}_u(n)$ are given in Eqs.\ (\ref{2.11}) and
(\ref{2.15a}), we consider a direct approach to calculating
$\tilde{K}(n_1,n_2)$ and
$\tilde{L}^{\prime}_u(n_1,n_2)$.  Denoting $K_{NS}(z)$,
$L_f(z)$ and
$L_u(z)$ collectively by $J(z)$, we determine the
double moments
\begin{eqnarray}
\tilde{J}(n_1, n_2) = \int ^1_0 dz z^{n_1-2}(1-z)^{n_2-1} J(z)
\label{3.15}
\end{eqnarray}
by expanding $(1-z)^{n_2-1}$ in powers of $z$ up to $O\left(
z^{n_2}\right)$, since $J(z)$ is small as $z \rightarrow 1$, and
the binomial coefficient $c_m$ decreases rapidly with $m$.  Thus
to a very good approximation we can write
\begin{eqnarray}
\tilde{J}(n_1, n_2) = \sum^{n_2}_{m = 0}(-1)^m {n_2 - 1
\choose m }\tilde{J}(n_1 + m)  ,
\label{3.16}
\end{eqnarray}
from which we can calculate $\tilde{K}(n_1, n_2)$ and
$\tilde{L}^{\prime}_{u}(n_1, n_2)$.

Having specified all the terms in Eqs.\ (\ref{3.5}) and
(\ref{3.7}), we can now calculate $\tilde{F}_{\pi^+}\left(n_1,
n_2 \right)$, which in turn are used in (\ref{3.8}) to determine
$\tilde{H}_{\pi^{\pm}}(n)$.  The inversion to
$H^{\prime}_{\pi^{\pm}}(x)$ [and then trivially to
$H_{\pi^{\pm}}(x)$] involves a method that has been discussed in
Ref.\
\cite{hy2} and is summarized in the Appendix.  We show our
results in Fig.\ 3.  The solid lines are from our calculations with
normalizations adjusted to fit the data \cite{b}, since the data are
for the dimensionful $Ed^3\sigma/dp^3$ at fixed
$p_T$ while our model is for the dimensionless $x dN/dx$,
integrated over all
$p_T$.  The shapes of the calculated inclusive distributions
involve no adjustable parameters.  Evidently, they agree very
well with the $x$ dependences of the data.  We regard this result
as evidence in support of the VRM.

\section{Pion Fragmentation:  $\pi^+ \rightarrow \pi^-$}

In this section we consider the non-diffractive inclusive
distribution of $\pi^+ \rightarrow \pi^-$ in the fragmentation
region of the pion beam.  By charge conjugate invariance it should
be the same as for $\pi^- \rightarrow \pi^+$, if there is no
contamination by target fragmentation.  Since there are no
valence quarks of the projectile that end up in the produced
pion, one naively would expect $dN/dx$ to be strongly
suppressed at high $x$.  Yet experimental data \cite{b} indicate
that its $x$ dependence is rather similar to that of $p
\rightarrow \pi^+$, in which a $u$ quark is shared between $p$
and $\pi^+$.  Thus simple fragmentation of a valence quark
or diquark in a string model cannot account for this similarity.
In the VRM because the valon distribution in the pion is
different from the ones in the proton, with the former giving
larger average momentum fraction of the valons than the latter,
large-$x$ $\pi^-$ in $\pi^+$ can arise from the
higher-momentum valons.  In this section we show this behavior
quantitatively.

For $\pi^+ \rightarrow \pi^-$ the invariant distribution
$H_{\pi^+ \rightarrow \pi^-}(x)$ has the same formal expression
as Eq.\ (\ref{3.1}) except that $F_{\pi}$ for $p$ fragmentation is
to be replaced by $F_{\pi^+ \rightarrow \pi^-}$.  In Fig.\ 4 we
show the schematic diagrams for $U\bar{D}$ valons going to
$d\bar{u}$ quarks.  The moments for $F_{\pi^+ \rightarrow
\pi^-}$ can then be written down by inspection
\begin{eqnarray}
\tilde{F}_{\pi^+ \rightarrow \pi^-}(n_1, n_2) =
\tilde{G}^{\pi}_{U\bar{D}}\left(n_1, n_2 \right)
\left[\tilde{L}^{\prime}_f(n_1)
\tilde{L}^{\prime}_f(n_2) + \tilde{L}^{\prime}_u(n_1)
\tilde{L}^{\prime}_u(n_2)
\right]\nonumber \\
+ \left[\tilde{G}^{\pi}_{U}\left(n_1
\right) + \tilde{G}^{\pi}_{\bar{D}}\left(n_2
\right) \right] \left\{
\tilde{L}^{\prime}_f(n_1,n_2)\tilde{L}_u(n_2)\right\}_{12}  .
\label{4.1}
\end{eqnarray}
We note that $\bar{u}$ is favored in $U$, and $d$ is favored in
$\bar{D}$.  Because we assume that the structure of valons is
universal in all host hadrons, the enhanced sea of a valon in the
pion is the same as that of a valon in the proton; thus
$\tilde{L}^{\prime}_{f,u}$ is the same as in the previous section.

Since $G^{\pi}_{U\bar{D}}\left(y_1, y_2 \right) = \delta \left(y_1
+y_2 - 1 \right)$ due to $\gamma = 0$ in Eq.\ (\ref{2.18}), we
have
\begin{eqnarray}
\tilde{G}^{\pi}_{U\bar{D}}\left(n_1,n_2\right) =
B\left(n_1,n_2\right) ,
\label{4.2}
\end{eqnarray}
\begin{eqnarray}
\tilde{G}^{\pi}_{U, \bar{D}}\left(n_i \right) = 1/n_i  .
\label{4.3}
\end{eqnarray}
The calculation of $\tilde{F}_{\pi^+ \rightarrow \pi^-}(n_1,
n_2)$ is therefore straightforward.  Using it in Eq.\ (\ref{3.8})
yields  $\tilde{H}_{\pi^+ \rightarrow \pi^-}(n)$, and by inversion
we get $H_{\pi^+ \rightarrow \pi^-}(x)$.

The result is shown in Fig.\ 5 by the solid line.  The dotted line is
the fit of the data given by the experimental paper \cite{b},
where the $x \geq 0.6$ points are excluded in order to avoid
contamination from resonance decay products.  Our result
agrees very well with the experimental parameterization of the
data by $ \sim (1 - x)^m$ where $m = 3.37 \pm 0.09$ for $p_T
= 0.3$ GeV/c at beam momentum 100 GeV/c.

The lack of $\tilde{K}$ in Eq. (\ref{4.1}) compared to (\ref{3.5})
and (\ref{3.7}), corresponding to no valence quarks, is
compensated by the fact that $\tilde{G}^{\pi}_{U\bar{D}}$ and
$\tilde{G}^{\pi}_{U, \bar{D}}$ in (\ref{4.2}) and (\ref{4.3}) are
not as damped at high $n_i$ as $\tilde{G}^p
_{UU}$, $\tilde{G}^p _{UD}$ and $\tilde{G}^p _{U,D}$ are in
(\ref{3.8a}), (\ref{3.8b}), (\ref{2.9}) and (\ref{2.10}),
respectively, for $\alpha$ and $\beta$ being as big as in
(\ref{2.5}).  That is, in $\pi^+$ we have harder valons and softer
$d$ and $\bar{u}$, while in $p$ we have softer valons and
harder $u$ and $\bar{d}$.

The normalization of the calculated distribution in Fig.\ 5 is
again adjusted to fit, since the data are for inclusive cross section
at fixed $p_T$.  Nevertheless, the agreement of the $x$
dependence with data is the second piece of support one can
infer for VRM.

\section{Kaon Fragmentation:  $K^+ \rightarrow \pi^{\pm}$}

In Sec. 2 we have discussed the form of the valon distributions in
a kaon with one undetermined parameter to be fixed.  Now we
consider kaon initiated reactions and determine that parameter
by fitting the inclusive distribution of $K^+ \rightarrow
\pi^{\pm}$.

From Eq.\ (\ref{2.20}) we have
\begin{eqnarray}
\tilde{G}^K_{U\bar{S}}\left(n_1,n_2\right) = B
\left(n_1 + a,n_2 + b\right)/B(a+1, b+1),
\label{5.1}
\end{eqnarray}
\begin{eqnarray}
\tilde{G}^K_{U}\left(n_1\right) = B
\left(n_1 + a, b+1\right)/B(a+1, b+1),
\label{5.2}
\end{eqnarray}
\begin{eqnarray}
\tilde{G}^K_{\bar{S}}\left(n_2\right) = B
\left(a+1, n_2 + b\right)/B(a+1, b+1),
\label{5.3}
\end{eqnarray}
where $a$ and $b$ are constrained by (\ref{2.23}).  In Fig.\ 6
we show the schematic diagrams for (a) $K^+ \rightarrow
\pi^+$ and (b) $K^+ \rightarrow \pi^-$.  It then follows that
\begin{eqnarray}
\tilde{F}_{K^+ \rightarrow \pi^+}(n_1, n_2) =
\tilde{G}^K_{U\bar{S}}\left(n_1, n_2 \right)
\left[\tilde{K}(n_1)
\tilde{L}^{\prime\prime}_u(n_2)
+ \tilde{L}^{\prime}_u(n_1)
\tilde{L}^{\prime\prime}_u(n_2) \right]\nonumber \\
+ \tilde{G}^K_{U}\left(n_1
\right)
\left\{\tilde{K}(n_1,n_2)\tilde{L}^{\prime}_u(n_2)\right\}_{12}\nonumber\\
+ \tilde{G}^K_{\bar{S}}\left(n_2\right)
\left\{\tilde{L}^{\prime\prime}_u(n_1,n_2)\tilde{L}^{\prime\prime}_u(n_2)\right\}_{12}
\label{5.3a}
\end{eqnarray}
where $\tilde{L}^{\prime\prime}_u(n_2)$ is the moment of the
enhanced, unfavored quark distribution in the $\bar{S}$ valon,
and will be discussed below.  Similarly, we have for $K^+
\rightarrow \pi^-$
\begin{eqnarray}
\tilde{F}_{K^+ \rightarrow \pi^-}(n_1, n_2) =
\tilde{G}^K_{U\bar{S}}\left(n_1, n_2 \right)
\left[\tilde{L}^{\prime}_f(n_1)
\tilde{L}^{\prime\prime}_u(n_2)
+ \tilde{L}^{\prime}_u(n_1)
\tilde{L}^{\prime\prime}_u(n_2) \right]\nonumber \\
+ \tilde{G}^K_{U}\left(n_1
\right)
\left\{\tilde{L}^{\prime}_u(n_1,n_2)\tilde{L}^{\prime}_f(n_2)\right\}_{12}\nonumber\\
+ \tilde{G}^K_{\bar{S}}\left(n_2\right)
\left\{\tilde{L}^{\prime\prime}_u(n_1,n_2)\tilde{L}^{\prime\prime}_u(n_2)\right\}_{12}.
\label{5.4}
\end{eqnarray}

For gluon conversion in the $\bar{S}$ valon we again consider
the saturation of the $u\bar{u}$ and $d\bar{d}$ sectors of the
sea, but none in the $s\bar{s}$ sector because of the higher
$s$-quark mass.  Since the non-strange sectors are both
unfavored, the momentum carried by the sea is not the same as
that in a non-strange valon.  Thus we write before gluon
conversion
\begin{eqnarray}
\tilde{K}_{NS}(2) + 4 \tilde{L}^S_u (2) + 2 \tilde{L}^S_s
(2) +  \tilde{L}^S_g (2) = 1 ,
\label{5.5}
\end{eqnarray}
where $\tilde{L}^S_i (i = u, s, g)$ refer to the $\bar{S}$
valon.  Equation (\ref{5.5}) differs from (\ref{3.9}) for a
$\bar{D}$ valon only in that the favored sector $d\bar{d}$ is
replaced by the unfavored sector $d\bar{d}$.  It means that
there is a redistribution of the momenta in the gluons and sea
quarks, which, we assume, takes the simple form
\begin{eqnarray}
\tilde{L}^S_i (2) = c \tilde{L}_i (2),
\label{5.6}
\end{eqnarray}
where $c$ is a constant.  After gluon conversion Eq.\ (\ref{5.5})
becomes
\begin{eqnarray}
\tilde{K}_{NS}(2) + 4 \tilde{L}^{\prime\prime}_u (2) + 2
\tilde{L}^S_s (2) = 1 .
\label{5.7}
\end{eqnarray}
If we define $\tilde{L}^{\prime\prime}_u (2) =
f_s\tilde{L}^S_s (2) $, we obtain
\begin{eqnarray}
f_s =1 + \tilde{L}^{S}_g (2)/4 \tilde{L}^S  _u(2) =
1+\tilde{L}_g (2)/4 \tilde{L}_u (2)
\label{5.8}
\end{eqnarray}
Using as before $\tilde{L}_u (2) = e^{-3.07}$ and $\tilde{L}_g
(2)= e^{-0.98}$, we get
\begin{eqnarray}
f_s = 3.02 .
\label{5.9}
\end{eqnarray}
To determine $\tilde{L}^{\prime\prime}_u (2)$ in terms of
$\tilde{L}_u (2)$ we need, in addition to $f_s$, the value of $c$
in Eq.\ (\ref{5.6}).  It follows from Eqs.\ (\ref{3.9}), (\ref{5.5})
and (\ref{5.6}) that
\begin{eqnarray}
c = 1 - {2\left[\tilde{L}_u (2) - \tilde{L}_f (2) \right]  \over  4
\tilde{L}_u (2) + 2 \tilde{L}_s (2) + \tilde{L}_g (2)}  .
\label{5.10}
\end{eqnarray}
Since $\tilde{L}_s (2) = e^{-4.21}$ from Table I, we get
\begin{eqnarray}
c = 0.9
\label{5.11}
\end{eqnarray}
and finally $\tilde{L}^{\prime\prime}_u (2) = f_s c \tilde{L}_u
(2) = 2.72 \tilde{L}_u (2)$.  Again, extending this
proportionality to all $n_2$, we have
\begin{eqnarray}
\tilde{L}^{\prime\prime}_u (n_2)= 2.72 \tilde{L}_u (n_2) .
\label{5.12}
\end{eqnarray}
We now have all the quantities in Eqs.\ (\ref{5.3}) and (\ref{5.4})
 to calculate $\tilde{F}_{K^+ \rightarrow \pi^{\pm}}(n_1, n_2)$.

There is one adjustable parameter in our calculation.  It is $a$ in
Eq.\ (\ref{2.20}), $b$ being constrained by (\ref{2.23}).  After
$\tilde{F}_{K^+ \rightarrow \pi^{\pm}}(n_1, n_2)$ are
determined, we use a formula similar to Eq.\ (\ref{3.8}), i.e.,
\begin{eqnarray}
\tilde{H}^{\prime}_{K^+ \rightarrow \pi^{\pm}}(n) =
\sum_{[n_i]} {n! \over n_1! n_2!} \tilde{F}_{K^+ \rightarrow
\pi^{\pm}}(n_1 + 2, n_2 + 2),
\label{5.13}
\end{eqnarray}
to calculate $\tilde{H}^{\prime}_{K^+ \rightarrow
\pi^{\pm}}(n)$.  Then by inversion to $H^{\prime}_{K^+ \rightarrow
\pi^{\pm}}(x)$ as before, we can finally obtain $H_{K^+
\rightarrow \pi^{\pm}}(x) = x ^{-3}\tilde{H}^{\prime}_{K^+
\rightarrow \pi^{\pm}}(x)$.  We adjust $a$ to fit the data,
which are shown in Fig.\ 7 for $p_T = 0.3$ GeV/c.  Again,
because the data are for fixed $p_T$, we cannot predict the
normalization, which is adjusted to fit.  The best values of $a$
and $b$ are
\begin{eqnarray}
a = 1.0 \quad , \qquad b = 2.0 .
\label{5.14}
\end{eqnarray}
The solid lines in Fig.\ 7 show the calculated results; the dotted
lines are the experimental fits using the form $(1 - x)^m$.
Although the fits are not perfect, they are acceptable in view of
the large error bars in the data.  What is noteworthy is that we
used only one normalization factor for both $\pi^+$ and $\pi^-$
production, and the calculated curves agree with the two sets of
data in their relative normalizations.  Thus the VRM has captured
the essence of $\pi^{\pm}$ production in $K^+$ initiated
reactions.

Since $a$ and $b$ are now known, we can exhibit the valon
distributions in a kaon.  In Fig.\ 8 we show $G^K_U(y)$ and
$G^K_{\bar{S}}(y)$.  Note how the strange valon has larger
momentum fraction than the nonstrange valon, on the average.
It is because of the harder $\bar{S}$ valon, though softer
non-strange quarks in it, that gives rise to the distributions
$H_{K^+ \rightarrow \pi^{\pm}}(x) $ that are harder
(decreasing more slowly with $x$) than either $\pi^+
\rightarrow \pi^-$ or even $p \rightarrow \pi^+$, which has a
shared valence quark.  This is a remarkable affirmation of the
importance of the structure of the hadron in the determination
of the inclusive distributions of its fragments.

\section{Pion Fragmentation:  $\pi^+ \rightarrow K^{\pm}$}

Since the valon distributions in the kaon have been determined
in the previous section, we now know the recombination
function for the formation of kaon.  Thus the calculation of pion
fragmentation into $K^{\pm}$ should be straightforward.

In Fig.\ 9 we show the diagrams for $\pi^+ \rightarrow
K^{\pm}$.  They are essentially the same as the ones in Fig.\ 6
for $K^+ \rightarrow \pi ^{\pm}$, except for the replacement of
$\bar{S}$ valon in the initial $K^+$ by $\bar{D}$ valon in the
initial $\pi^+$, and the replacement of the appropriate
non-strange $q$ and $\bar{q}$ quarks in $\pi^{\pm}$ by the
$s$ and $\bar{s}$ quarks in $K^{\pm}$.  The corresponding
equations are therefore similar to Eqs.\ (\ref{5.3}) and
(\ref{5.4})
\begin{eqnarray}
\tilde{F}_{\pi^+ \rightarrow K^+}(n_1, n_2) =
\tilde{G}^{\pi}_{U\bar{D}}\left(n_1, n_2 \right)
\left[\tilde{K}(n_1)\tilde{L}_s(n_2)+ \tilde{L}_s(n_1)
\tilde{L}^{\prime}_u(n_2) \right]\nonumber \\
+ \tilde{G}^{\pi}_{U}\left(n_1
\right)
\left\{\tilde{K}(n_1,n_2)\tilde{L}_s(n_2)\right\}_{12}\nonumber\\
+ \tilde{G}^{\pi}_{\bar{D}}\left(n_2\right)
\left\{\tilde{L}^{\prime}_u(n_1,n_2)\tilde{L}_s(n_2)\right\}_{12}
\label{6.1}
\end{eqnarray}
\begin{eqnarray}
\tilde{F}_{\pi^+ \rightarrow K^-}(n_1, n_2) =
\tilde{G}^{\pi}_{U\bar{D}}\left(n_1, n_2 \right)
\left[ \tilde{L}^{\prime}_f(n_1)\tilde{L}_s(n_2)
+\tilde{L}_s(n_2) \tilde{L}^{\prime}_u(n_1)
 \right]\nonumber \\
+ \tilde{G}^{\pi}_{U}\left(n_1
\right)
\left\{\tilde{L}^{\prime}_f(n_1,n_2)\tilde{L}_s(n_2)\right\}_{12}\nonumber\\
+ \tilde{G}^{\pi}_{\bar{D}}\left(n_2\right)
\left\{\tilde{L}^{\prime}_u(n_1,n_2)\tilde{L}_s(n_2)\right\}_{12}
.
\label{6.2}
\end{eqnarray}
Note that $\tilde{L}_s(n_i)$ is not enhanced because, as before,
gluon conversion saturates the non-strange sectors of the sea.
The other moments are as given in Sec.\ 4.

Since the recombination function $R_K$ is now different from
$R_{\pi}$, the inclusive distribution for $K^{\pm}$ production
is
\begin{eqnarray}
x {dN_K \over dx} &=& H_K (x) = \int {dx_1 \over x_1} {dx_2
\over x_2} F_{\pi \rightarrow K} \left(x_1,
x_2\right)R_K\left(x_1, x_2, x\right)\nonumber \\
&=& g_Kx^{-4} \int dx_1 dx_2  x_1x_2^2 F_{\pi \rightarrow K}
\left(x_1, x_2\right)\delta \left(x_1 +x_2-x\right)
\label{6.3}
\end{eqnarray}
where Eqs.\ (\ref{2.25}) and (\ref{5.14}) have been used.  If we
define
\begin{eqnarray}
H^{\prime}_K(x) = x^6 H_K(x) ,
\label{6.4}
\end{eqnarray}
then the binomial expansion of $\left(x_1 + x_2 \right)^n$ leads
us to the moments
\begin{eqnarray}
\tilde{H}^{\prime}_K(n) = \int^1_0 dx x^{n-2}H^{\prime}_K(x) =
g_K \sum_{[n_i]}{n! \over n_1! n_2!} \tilde{F}_{\pi \rightarrow
K}\left(n_1 + 3, n_2 + 4 \right) ,
\label{6.5}
\end{eqnarray}
similar, but not identical, to Eq.\ (\ref{3.8}). In view of
Eqs.\ (\ref{6.1}) and (\ref{6.2}),
$\tilde{H}^{\prime}_{K^{\pm}}(n)$ can now be calculated
without any free parameters.

Using the same procedure as before to obtain
$H_{K^{\pm}}(x)$, we can determine the
inclusive distributions for $\pi^+ \rightarrow K^{\pm}$. In Fig.\ 10
we show our results compared to the data \cite{b} at 100 GeV/c
and
$p_T=0.3$ GeV/c. Again, the normalization is adjusted to fit, but
only one normalization factor for  both curves. The agreement
between theory and experiment is very good,  considering that no
free parameter is used except for the overall normalization. The
excess for $x>0.6$ can be attributed to the decay of $K^*$ whose
$K$ product  would have higher $x$. The relative normalization of
$K^+$ to $K^-$ is well reproduced by VRM.

\section{Conclusion}

We have considered the production of mesons in the
fragmentation regions of incident proton and mesons in
low-$p_T$ collisions. The data on $p\rightarrow \pi^{\pm}$,
$\pi^+ \rightarrow \pi^-$,
$K^+\rightarrow \pi^{\pm}$, and $\pi^+ \rightarrow K^{\pm}$
have all been satisfactorily reproduced by the calculated results in
the VRM. There is essentially only one free parameter that is
connected with the valon distribution in the kaon. All other
parameters specifying the structure of the proton and pion have
been determined independently by fitting the data on hard
processes. It is self-evident that by successfully reproducing the
inclusive distributions of all the above reactions we have met the
test of charge asymmetry in $\pi^+p$ collisions.

Apart from the details of the VRM and of the data in the
fragmentation region, the fundamental themes that this work has
affirmed are that the hadron structures are important and that
hadronization proceeds through recombination. The valon model
effectively describes the hadron structure and the valon
distributions provide the probability functions for recombination.
It is conceivable that two valons may be regarded as a diquark
whose fragmentation yields the produced mesons. Since the two
valons are spatially distributed objects with non-vanishing relative
momentum, to describe their fragmentation by a fragmentation
function adapted from jet physics seems hard to justify. The VRM,
on the other hand, is related more closely to the parton model
and provides a natural s-channel description of the fragmentation
process (from the point of view of the hadron) in terms of
recombination (from the point of view of the quarks and
antiquarks).

We have not considered the production of baryons in this paper.
However, the non-diffractive production of nucleons in $pA$
collisions has already been treated in the VRM \cite{hy2}. The
production of strange particles is worthy of further attention. Our
consideration of $\pi^+ \rightarrow K^{\pm}$ is only a beginning,
which shows that in hadronic collisions gluons do not convert to
$s\bar{s}$ as effectively as to $u\bar{u}$ and $d\bar{d}$. That
situation must change in going from hadronic to nuclear collisions
due to Pauli blocking in the non-strange sector. It seems that the
VRM provides a natural framework in which to investigate the
transition from strangeness suppression to strangeness
enhancement.

\section*{Acknowledgment} One of us (RCH) is grateful to Prof. A.
Capella for stimulating discussions that have led to a challenge for
us to show how well the recombination model works.  This work was
supported, in part,  by the U.\ S.\ Department of Energy under
Grant No. DE-FG03-96ER40972.

\newpage
\appendix
\section{Appendix}
We summarize here the method of inversion from the moments to
the $x$-distribution function, originally proposed in Ref.\
\cite{hy2}.  Instead of making the inverse Mellin transform, which
involves a complex contour, we exploit the orthogonality of the
Legendre polynomials.  First, we shift the variable to the interval
$0 \leq x \leq 1$, and define
\begin{eqnarray}
g_{\ell}(x) = P_{\ell}(2x - 1)
\label{A1}
\end{eqnarray}
so that
\begin{eqnarray}
\int^1_0 dx g_{\ell}(x)g_m(x) = { 1 \over 2\ell +1}
\delta_{\ell m}.
\label{A2}
\end{eqnarray}
If we expand the distribution $H'(x)$ in terms of
$g_{\ell}(x)$
\begin{eqnarray}
H'(x) = \sum^{\infty}_{\ell = 0} (2\ell
+1)h_{\ell}g_{\ell}(x),
\label{A3}
\end{eqnarray}
then the inverse is
\begin{eqnarray}
h_{\ell} = \int^1_0 dx H'(x)g_{\ell}(x).
\label{A4}
\end{eqnarray}
These $h_{\ell}$ can be expressed in terms of the moments
$H'(n)$ if we express $g_{\ell}(x)$ as a power series in $x$
\begin{eqnarray}
g_{\ell}(x) = \sum^{\ell}_{i = 0}a^i_{\ell}x^i ,
\label{A5}
\end{eqnarray}
where $a^i_{\ell}$ are known from the properties of
$P_{\ell}(z)$.  Thus from Eq.\ (\ref{A4}) we have
\begin{eqnarray}
h_{\ell} = \sum^{\ell}_{i = 0}a^i_{\ell} \tilde{H}'(i+2) ,
\label{A6}
\end{eqnarray}
where $\tilde{H}'(n)$ is defined in Eq.\ (\ref{3.8}).  It is
now clear that our theoretical results in $\tilde{H}'(n)$ can
be transformed to $H'(x)$ through Eqs.\
(\ref{A3}) and (\ref{A6}) once we have the coefficients
$a^i_{\ell}$.  Furthermore, if $\tilde{H}'(n)$ becomes
unimportant for $n > N$, then the sum in Eq.\ (\ref{A3})
can terminate at $N$.

To determine $a^i_{\ell}$, we make use of the recursion
formula
\begin{eqnarray}
(\ell + 1) P_{\ell + 1}(z) = (2\ell + 1) z P_{\ell}(z) -
\ell P_{\ell - 1}(z)
\label{A7}
\end{eqnarray}
to infer through Eqs.\ (\ref{A1}) and (\ref{A5})
\begin{eqnarray}
a^0_{\ell} = -{1 \over \ell}[(2 \ell-1) a^0_{\ell -1} +
(\ell-1) a^0_{\ell -2}],
\label{A8}
\end{eqnarray}
\begin{eqnarray}
a^i_{\ell} = -{1 \over \ell}[(2 \ell-1) (a^i_{\ell -1} -
2a^{i-1}_{\ell -1} )+ (\ell-1) a^i_{\ell -2}],
\label{A9}
\end{eqnarray}
where $\ell \geq 2$, and $1 \leq i \leq \ell$.  For  $\ell < 2$, we have
\begin{eqnarray}
a^0_0 = 1, \qquad a^0_1 = -1, \qquad a^1_1 = 2 .
\label{A10}
\end{eqnarray}
With these we can generate all $a^i_{\ell}$, so $h_{\ell}$
can be directly computed.  The use of (\ref{A3}) then yields
$H^{\prime}(x)$.

We have found that this method can give very accurate result in
inverting $\tilde{H}^{\prime}(n)$ to $H^{\prime}(x)$ for $N$
roughly between 8 to 10, depending on how rapidly
$\tilde{H}^{\prime}(n)$ decreases with $n$.

\begin{figure}[tbh]
\includegraphics[width=0.45\textwidth]{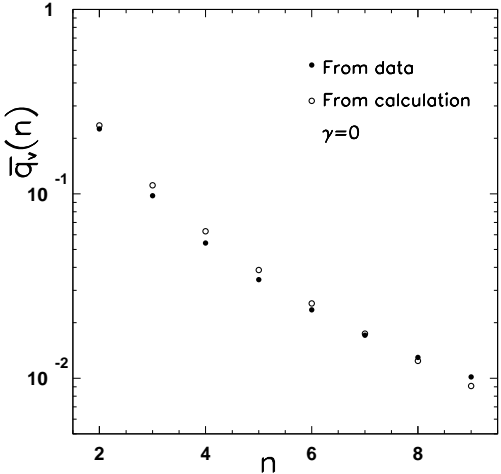}
\caption{
Moments of the valence quark distribution in a pion,
$\tilde q_v(n)$, where the parameter $\gamma$ is chosen to fit the
moments of distribution determined from the experimental data
\cite{sut}.}
\end{figure}

\begin{figure}[tbh]
\includegraphics[width=0.45\textwidth]{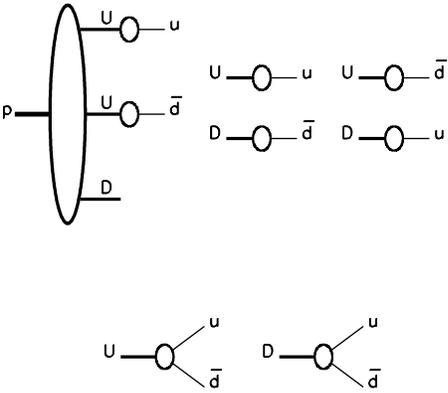}
\caption{
 Schematic diagrams in the VRM for the production
of $\pi^+$ and $\pi^-$ in the proton fragmentation region; only
the $u\bar{d}$ and $d\bar{u}$ states are shown.}
\end{figure}

\begin{figure}[tbh]
\includegraphics[width=0.45\textwidth]{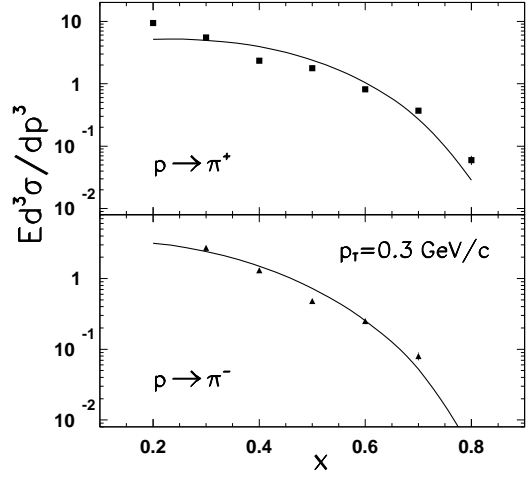}
\caption{
Inclusive distributions of $p \rightarrow
\pi^+$ and $\pi^-$.  The data are from Ref.\ \cite{b} at
$P_L = 100$ GeV/c and $p_T = 0.3$ GeV/c.}
\end{figure}

\begin{figure}[tbh]
\includegraphics[width=0.45\textwidth]{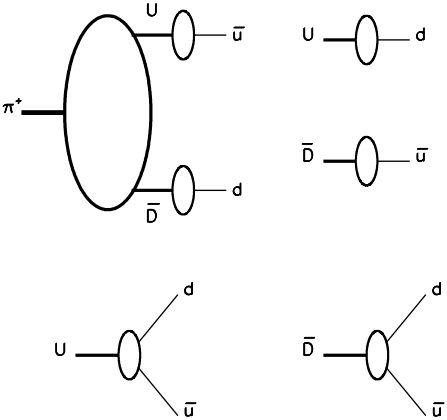}
\caption{
Schematic diagrams in the VRM for
$\pi^+ \rightarrow d\bar{u}$.}
\end{figure}

\begin{figure}[tbh]
\includegraphics[width=0.45\textwidth]{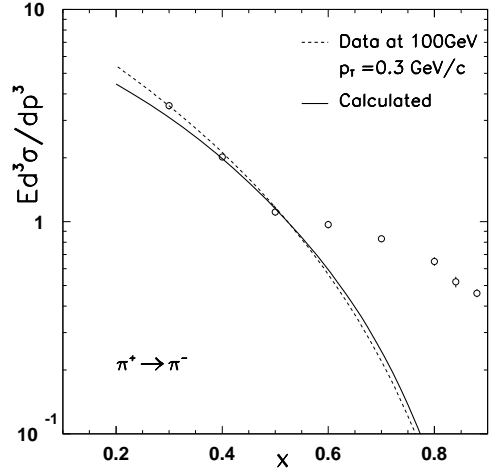}
\caption{
 Inclusive distributions of $\pi^+\rightarrow
\pi^-$.  Data are from Ref.\ \cite{b}.  The dotted line is the
experimental fit; the solid line is the theoretical result in the VRM.}
\end{figure}

\begin{figure}[tbh]
\includegraphics[width=0.45\textwidth]{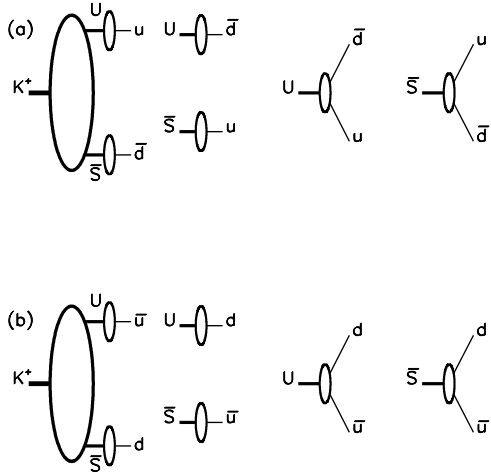}
\caption{
 Schematic diagrams in the VRM for (a) $K^+
\rightarrow u\bar{d}$, and (b) $K^+ \rightarrow d\bar{u}$.}
\end{figure}

\begin{figure}[tbh]
\includegraphics[width=0.45\textwidth]{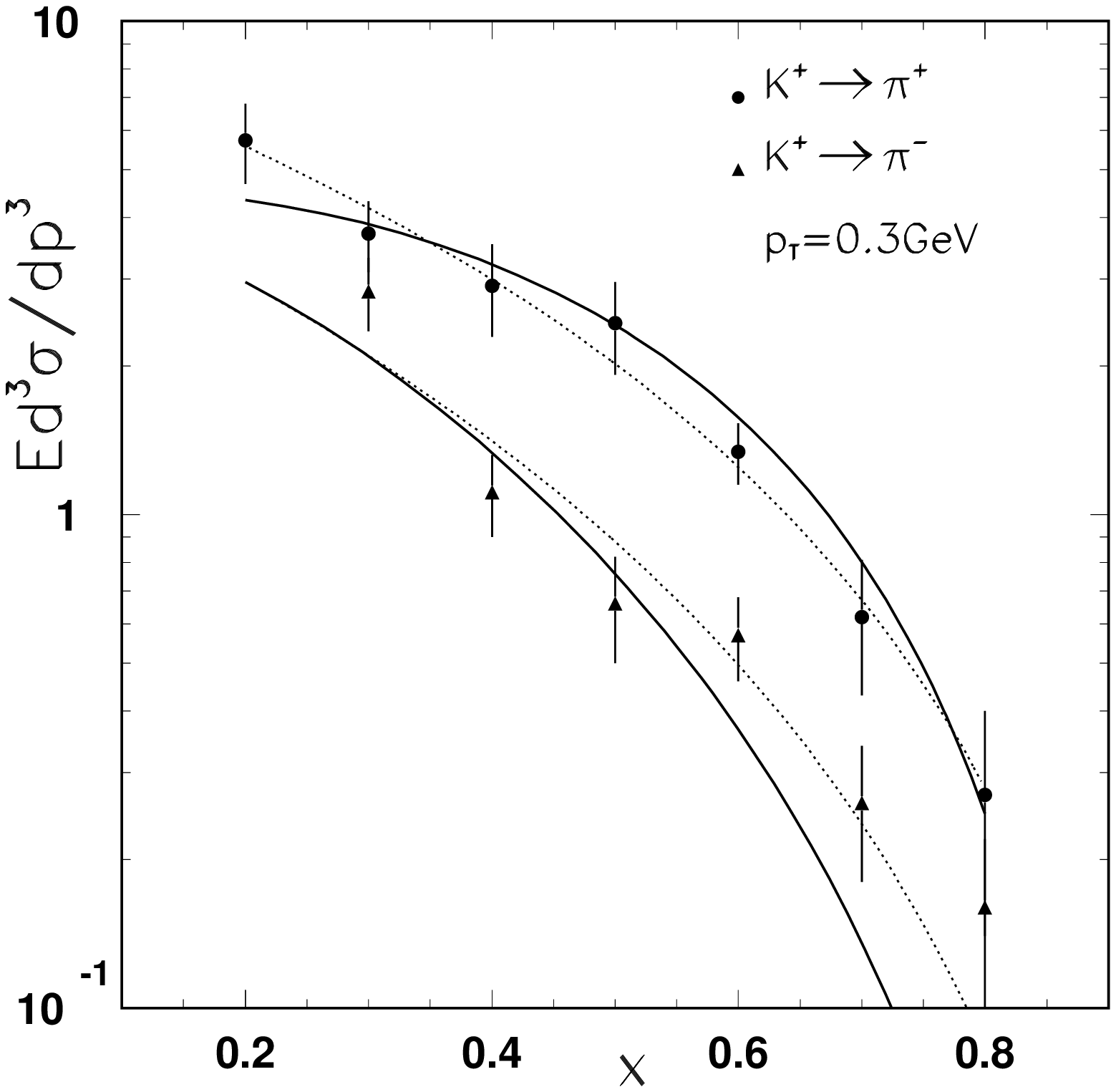}
\caption{
 Inclusive distributions of $K^+
\rightarrow \pi^+$ and  $K^+ \rightarrow \pi^-$.  The data are
from Ref.\ \cite{b} $P_L = 100$ GeV/c and $p_T = 0.3$ GeV/c.
The dotted lines are the experimental fits; the solid lines are the
theoretical results in the VRM.}
\end{figure}

\begin{figure}[tbh]
\includegraphics[width=0.45\textwidth]{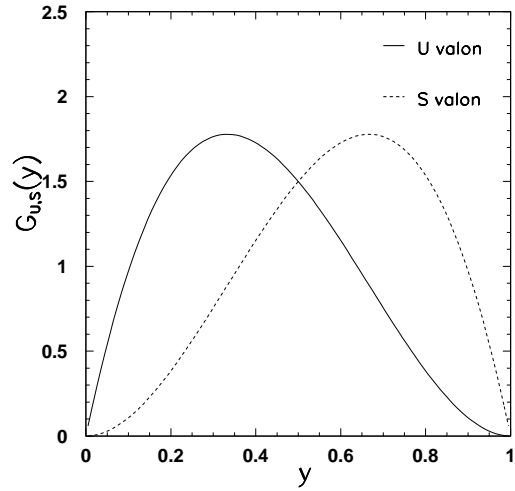}
\caption{
 The momentum-fraction distributions of the
valons in a kaon:  (a) $U$ valon in solid lines, and (b) $\bar{S}$
valon in dotted line.}
\end{figure}

\begin{figure}[tbh]
\includegraphics[width=0.45\textwidth]{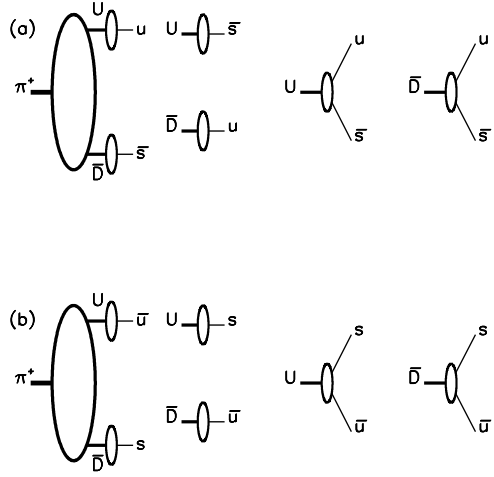}
\caption{
 Schematic diagrams in the VRM for (a)
$\pi^+ \rightarrow u\bar{s}$, and (b) $\pi^+ \rightarrow
s\bar{u}$.}
\end{figure}

\begin{figure}[tbh]
\includegraphics[width=0.45\textwidth]{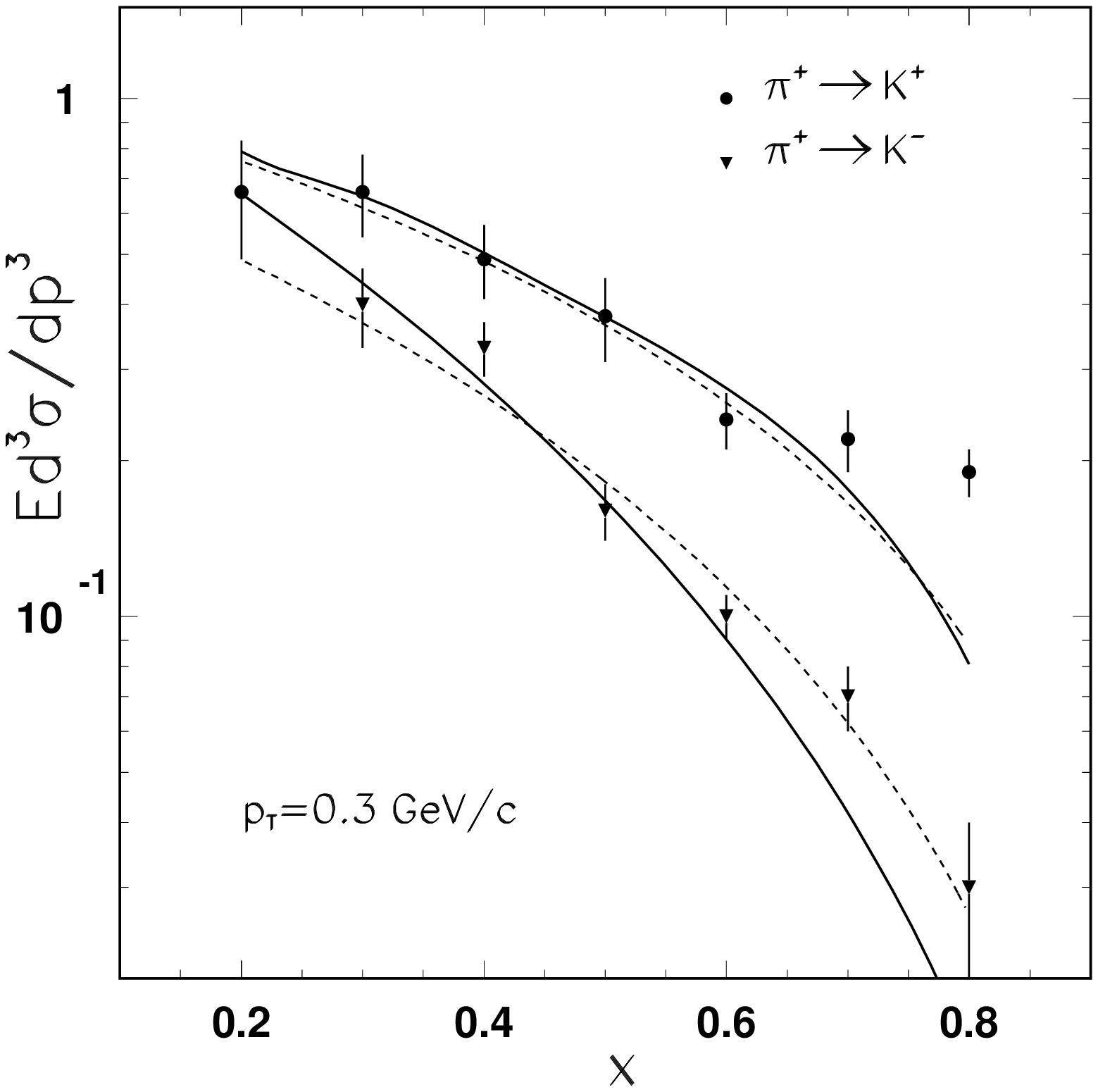}
\caption{
Inclusive distributions of $\pi^+ \rightarrow
K^+$, and (b) $\pi^+ \rightarrow K^-$. The data are
from Ref.\ \cite{b} $P_L = 100$ GeV/c and $p_T = 0.3$ GeV/c.
The dotted lines are the experimental fits; the solid lines are the
theoretical results in the VRM.}
\end{figure}


\begin{thebibliography}{99}

\bibitem{vw} {\it Partons in Soft-Hadronic Processes}, edited by
R.\ T.  Van de Walle (World Scientific, Singapore, 1981).

\bibitem{fk}K.\ Fia\l kowski and  W.\ Kittel, Rep.\ on Progress in
Physics {\bf 46}, 1283 (1983).

\bibitem{dpm}
A.\ Capella, U.\ Sukhatme, C.-I.\ Tan and J.\ Tran Thanh Van,
Phys.\ Report {\bf 236}, 225 (1994).

\bibitem{lu}
B.\ Andersson, G.\ Gustafson, and C.\ Peterson,
Phys.\ Lett.\ {\bf 69B}, 221 (1977); {\bf 71B}, 337 (1977); B.\
Andersson, G.\ Gustafson, G.\ Ingelman, and T.\ Sj\"{o}strand,
Phys. Rep. {\bf 97}, 33 (1983).

\bibitem{dh}
K.\ P.\ Das and R.\ C.\ Hwa, Phys.\ Lett.\ {\bf
68B}, 459 (1977).

\bibitem{val}
R.\ C.\ Hwa, Phys.\ Rev.\ D{\bf 22}, 1593 (1980).

\bibitem{cteq}
H.\ L.\ Lai, {\it et al.}, CTEQ Collaboration,
 Phys.\ Rev.\ D{\bf 51}, 4763 (1995).

\bibitem{hz} R.\ C.\ Hwa, Phys.\ Rev.\ D{\bf 22},
759 (1980); R.\ C.\ Hwa, and M.\ S.\ Zahir, Phys.\ Rev.\
D{\bf 23}, 2539 (1981).

\bibitem{hy}R.\ C.\ Hwa and C.\ B.\ Yang, hep-ph/0202140
(submitted to Phys.\ Rev.\ C).

\bibitem{wo}W.\ Ochs, Nucl.\ Phys.\ {\bf B118}, 397 (1977).

\bibitem{fig}Figure 4.8 in \cite{dpm}.

\bibitem{b}A.\ E.\ Brenner {\it et al.}, Phys.\ Rev.\
D{\bf 26}, 1497 (1982).

\bibitem{hy2}R.\ C.\ Hwa, and C.\ B.\ Yang,
Phys.\ Rev.\ C {\bf 65}, 034905 (2002).

\bibitem{cteq4}CTEQ4LQ\\
http://zebu.uoregon.edu/$\sim$parton/partongraph.html

\bibitem{sut}P.\ J.\ Sutton, A.\ D.\ Martin, R.\ G.\ Roberts,
and W.\ J.\ Stirling, Phys.\ Rev.\ D{\bf 45}, 2349 (1992).
\end{thebibliography}
\end{document}